\documentstyle[12pt,epsfig]{article}
\def\lbldef#1#2{\expandafter\gdef\csname #1\endcsname {#2}}

\def\href#1#2{#2}  

\begin{document}
\baselineskip=15.5pt
\pagestyle{plain}
\setcounter{page}{1}
\renewcommand{\thefootnote}{\fnsymbol{footnote}}
\begin{titlepage}

\vspace*{-3cm} \hfill 
\raggedleft{CERN-TH/99-72\\ hep-th/9903186}

\begin{center}

\vskip 4cm

{\Large {\bf Branes at Singularities in Type 0 String Theory
}}

\vskip 1.5cm

{\large Mohsen Alishahiha, Andreas Brandhuber and Yaron Oz}
\vskip 0.5cm

Theory Division, CERN,
CH-1211 Geneva 23, Switzerland 

\vskip 2cm

\begin{abstract}
We consider Type 0B D3-branes placed 
at conical singularities and analyze  in detail the conifold singularity.
We study the non supersymmetric gauge theories on their worldvolume and 
their conjectured dual gravity descriptions.
In the ultraviolet the solutions exhibit a logarithmic running
of the gauge coupling. In the infrared we find confining solutions and IR fixed points. 

\end{abstract}
\end{center}
\end{titlepage}

\newpage
\def\ads{{\it AdS}}
\def\adsp{{\it AdS}$_{p+2}$}
\def\cft{{\it CFT}}

\newcommand{\beq}{\begin{equation}}
\newcommand{\eeq}{\end{equation}}
\newcommand{\ber}{\begin{eqnarray}}
\newcommand{\cN}{{\cal N}}
\newcommand{\cC}{{\cal C}}
\newcommand{\eer}{\end{eqnarray}}
\newcommand{\cD}{{\Delta}}
\newcommand{\be}{\begin{equation}}
\newcommand{\ee}{\end{equation}}
\newcommand{\eel}[1]{\label{#1}\end{equation}}
\newcommand{\bea}{\begin{eqnarray}}
\newcommand{\eea}{\end{eqnarray}}
\newcommand{\eeal}[1]{\label{#1}\end{eqnarray}}
\newcommand{\baq}{\begin{equation}\begin{array}{rcl}}
\newcommand{\eaq}{\end{aryray}\end{equation}}
\newcommand{\eaql}[1]{\end{array}\label{#1}\end{equation}}
\newcommand{\beac}{\begin{equation}\begin{array}{rcl}}
\newcommand{\eeacn}[1]{\end{array}\label{#1}\end{equation}}
\newcommand{\ba}{\begin{array}}
\newcommand{\ea}{\end{array}}
\newcommand{\non}{\nonumber \\}
\newcommand{\equ}[1]{(\ref{#1})}

\renewcommand{\a}{\alpha}
\renewcommand{\b}{\beta}
\newcommand{\g}{\gamma}           \newcommand{\G}{\Gamma}
\renewcommand{\d}{\delta}         \newcommand{\D}{\Delta}
\newcommand{\e}{\varepsilon}
\renewcommand{\k}{\kappa}
\newcommand{\la}{\lambda}        \newcommand{\La}{\Lambda}
\newcommand{\m}{\mu}
\newcommand{\n}{\nu}
\newcommand{\om}{\omega}         \newcommand{\OM}{\Omega}
\newcommand{\p}{\psi}             \newcommand{\PS}{\Psi}
\renewcommand{\r}{\rho}
\newcommand{\s}{\sigma}           \renewcommand{\S}{\Sigma}
\renewcommand{\th}{\theta}         \newcommand{\T}{\Theta}
\newcommand{\f}{{\phi}}           \newcommand{\F}{{\Phi}}
\newcommand{\vf}{{\varphi}}
\newcommand{\y}{{\upsilon}}       \newcommand{\Y}{{\Upsilon}}
\newcommand{\z}{\zeta}
\newcommand{\ggN}{g_{YM}^2 N}
\renewcommand{\AA}{{\mathcal A}}
\newcommand{\BB}{{\mathcal B}}
\newcommand{\CC}{{\mathcal C}}
\newcommand{\DD}{{\mathcal D}}
\newcommand{\EE}{{\mathcal E}}
\newcommand{\FF}{{\mathcal F}}
\newcommand{\GG}{{\mathcal G}}
\newcommand{\HH}{{\mathcal H}}
\newcommand{\II}{{\mathcal I}}
\newcommand{\JJ}{{\mathcal J}}
\newcommand{\KK}{{\mathcal K}}
\newcommand{\LL}{{\mathcal L}}
\newcommand{\MM}{{\mathcal M}}
\newcommand{\NN}{{\mathcal N}}
\newcommand{\OO}{{\mathcal O}}
\newcommand{\PP}{{\mathcal P}}
\newcommand{\QQ}{{\mathcal Q}}
\renewcommand{\SS}{{\mathcal S}}
\newcommand{\RR}{{\mathcal R}}
\newcommand{\UU}{{\mathcal U}}
\newcommand{\VV}{{\mathcal V}}
\newcommand{\WW}{{\mathcal W}}
\newcommand{\XX}{{\mathcal X}}
\newcommand{\YY}{{\mathcal Y}}
\newcommand{\ZZ}{{\mathcal Z}}

\newcommand{\w}{Schwarzschild $\:$}
\newcommand{\va}{\varphi}
\newcommand{\de}{\Delta}
\newcommand{\x}{\tilde{x}}

\section{Introduction}

The Type 0 string theories have worldsheet supersymmetry but no space-time 
supersymmetry as a consequence of 
non-chiral GSO projection \cite{Dixon:1986ba,Seiberg:1986ss}.
Consider two types of such string theories, Type 0A and Type 0B.
They do not have space-time fermions in their spectra. 
Nevertheless, they have a modular invariant partition function.
The bosonic fields of these theories are as those of the supersymmetric
Type IIA and Type IIB string theories, with a doubled set of Ramond-Ramond
fields. 
In addition they contain a tachyon field $T$.

Type 0 theories have D-branes. As in the Type II case, 
one can consider the gauge theories 
on the worldvolume of $N$ such branes.
These theories do not contain an open string tachyon \cite{Polyakov:1998ju}.
Moreover, the bulk tachyon can condense due to its coupling to 
the Ramond-Ramond fields.

One example studied in 
\cite{Klebanov:1998db,Klebanov:1998af,Minahan:1998gm,Minahan:1998af} 
is the theory on D3 branes in Type 0B theory. 
Since there is a doubled set of RR 4-form fields in
Type 0B string theory, the D3 branes can carry two charges,
electric and magnetic.
The worldvolume theory of $N$ flat electric 
D3 branes is a $U(N)$ gauge theory with six scalars
in the adjoint representation of the gauge group.

Asymptotic solutions of 
a proposed dual gravity background were constructed in \cite{Klebanov:1998db}.
At large radial coordinate the tachyon is constant and  one finds a metric 
of the form $AdS_5 \times S^5$ with vanishing coupling which was interpreted as
a UV fixed point. 
The solution exhibits a logarithmic running in qualitative agreement
with the asymptotic freedom property of the field theory \cite{Klebanov:1998af,Minahan:1998gm}.
Moreover the sign of the second coefficient of the $\beta$-function
was found to be in agreement with field theory \cite{Klebanov:1998af}.

At small radial coordinate the tachyon vanishes and one finds again a solution
of the form $AdS_5 \times S^5$ with 
infinite coupling, which was interpreted as 
a strong coupling IR fixed point \cite{Klebanov:1998af}.
The complete gravity solution describing the flow from the UV fixed point 
to the IR fixed point has not been constructed yet.

Generically one expects
the gauge theory to have different phases parametrized by the possible  
couplings.
The IR fixed point should occur as a particular tuning of the 
couplings.
Indeed, other solutions at small radial coordinate were constructed
in \cite{Minahan:1998af} that  exhibit confinement and mass gap.
Moreover they were argued to be more generic than the IR fixed point solution.
  
One can construct other non supersymmetric
theories by placing the Type 0B D3 branes at singularities.
In this paper we will pursue this direction.
We will mostly consider the theories obtained by placing Type 0B D3 branes
at a conifold singularity, but we will also discuss general features 
applicable to other singularities as well.
The paper is organized as follows.
In the next section we will comment on the phase structure of the worldvolume
theory of $N$ flat electric D3 branes of Type 0B string theory and the dual 
gravity description.
The discussion will apply as well to branes at singularities which will 
be studied later.
In section 3 we will place Type 0B D3 branes at a conifold singularity.
We will consider electric branes, magnetic branes and self-dual branes.
We will discuss the worldvolume field theory and construct conjectured dual 
gravity
descriptions. These exhibit UV freedom and a logarithmic running of 
the gauge coupling.
However, we will see that the sign of the two-loop $\beta$-function 
coefficient is not captured correctly by gravity.
In the IR we find gravity solutions that correspond to  fixed points
as well as solutions that exhibit confinement.
In section 4 we will discuss Type 0B D3 branes placed in general 
singular spaces.     
Section 5 is devoted to a discussion on the RG flow from the UV to 
the IR in the gravity description.

\section{Comments on phase structure}

In the following we make some comments that will be relevant later too,
 on the 
theory of $N$ flat electric D3 branes of Type 0B string theory.
The theory is a
non supersymmetric  
$SU(N)$ gauge theory with six real scalars $X^i,i=1,...,6$
in the adjoint representation of the gauge group.
The classical action is derived by a dimensional reduction of the pure
$SU(N)$ gauge theory action in ten dimensions.
The six scalars are the components of the gauge fields in the reduced
dimensions.

The theory has classically $SO(6)$ global symmetry that rotates the 
six scalars.
This allows three possible parameters:
a gauge coupling $g_{YM}$, a mass parameter for the scalars $m$
and a scalar quartic potential coupling $g$.
In the classical action, the mass parameter
is zero and $g$ is fixed in terms of $g_{YM}$.
Quantum mechanically, the three parameters 
are corrected differently and can take 
independent values.
The theory has a phase diagram depending on these three parameters.
Generically we expect to see in the diagram Coulomb-like (Higgs) phases, 
confinement phases
and maybe non trivial RG fixed points arising from particular tuning of 
the parameters.
For instance, such a picture has been seen on the lattice for $SU(2)$ 
gauge theory with one scalar in the
adjoint representation \cite{Baier:1986ts, Baier:1987ts, Baier:1988tr}. One starts with the theory with 
three bare parameters, the gauge coupling,  a mass parameter for the scalar and
a scalar quartic potential coupling.
The phase diagram is constructed
by tuning the bare parameters to different values and taking the continuum
limit. The case with six scalars in the adjoint representation has not 
been analysed on the lattice yet.

We may expect, like in the case with one scalar, that the 
vacuum expectation value for the scalars
will vanish
in the confining phase and be non zero in the Higgs phase. 
A non trivial fixed point may appear on the border of these phases at 
zero vacuum expectation value for the scalars.
Since the gravity description that we consider corresponds to D3 branes on top of each other,
the vacuum expectation value for the scalars is zero.
We expect to see then a confining phase and maybe IR fixed points.

The electric D3 branes theory is conjectured to have a dual string description.
There are two approximations of the dual description that can be taken.
When the effective string coupling $e^{\phi}$
is small string loops are suppressed
and we can use the classical string description.
When the $\alpha'$ corrections are small the classical
gravity description is applicable.

The gauge theory is asymptotically free. At short distances
the theory is weakly coupled and we expect the dual gravity 
description not to be valid.
This can happen for instance if the gravity background is singular 
as in the models 
based on heating up theories in one higher dimension 
\cite{Witten:1998as,Csaki:1999ln}.
In the case at hand the signal for the gravity description being 
non adequate
is that the $\alpha'$ corrections are of the same order  as the 
leading order gravity contributions \cite{Klebanov:1998af}.
For large $N$ the curvature in string units is small but $\alpha'$ corrections
associated with the Weyl tensor are not suppressed. 
The effective string coupling is small and therefor we expect the classical
string description to be applicable.
This still has an advantage over the finite temperature case in 
that we can see
qualitatively the asymptotic freedom property already in the leading order
gravity description.
In \cite{Klebanov:1998af} it was argued that the gravity  description at large
radial coordinate captures 
the right sign of the one and two loop $\beta$ function coefficients.

The gravity description cannot capture quantitatively the renormalization
group flow from the UV to the IR.
For the electric D3 branes there is a gravity solution that
indicates an IR fixed point at infinite coupling.
While the $\alpha'$ corrections are suppressed in this region, the dilaton 
is large
and classical string theory
is not sufficient
to study the fixed point theory.
The gravity solution at all energy scales $u$ has not been constructed yet.
Other gravity solutions with a confining 
behaviour in the IR were found in \cite{Minahan:1998af} and will be 
discussed in the following.

The ansatz for the background metric (in string frame) is 
\cite{Klebanov:1998db}
\beq
ds =  e^{\frac{1}{2}\phi}\left(e^{\frac{1}{2} \xi - 5\eta} d\rho^2 +
e^{-\frac{1}{2} \xi} dx^{\mu} dx^{\mu} + 
e^{\frac{1}{2} \xi - \eta}d\Omega_5^2 \right) \ ,
\label{metric}
\eeq
where $\phi,\xi,\eta$ are functions of the radial coordinate 
$\rho \sim 1/u^4$ and $\phi$ denotes the dilaton field.
The tachyon field $T$ induces a change in the dilaton field in the radial
coordinate $\rho$. This corresponds to the
running of the gauge coupling in the gauge theory.
Different forms of the tachyon fields as a function of $\rho$ induce
different RG trajectories from the UV to the IR.
Changing the form of the tachyon field and as a consequence
of the equations of motion changing the form of $\phi,\xi,\eta$
is analogus to changing the parameters $g_{YM},m,g$.
In analogy with the lattice picture, 
we expect the gravity description to exhibit
different phases parametrized by the form of $\phi,\xi,\eta$ and $T$.

The mapping of parameters between the field theory and gravity cannot be made 
quantitatively precise. 
One can see, however, that $\xi,\eta$ together with $\phi$
control the size of the $S^5$ part of the metric (\ref{metric}).
One expects that the signal for the scalars in the adjoint
representation being massive is the shrinking
of the five sphere to zero size.

Consider the following  large $\rho$
behaviour of $\phi,\xi,\eta$ \cite{Minahan:1998af}
\ber
\phi & \sim & \alpha_0 \rho + \phi_0 \nonumber\\
\xi & \sim & \alpha_1 \rho + \xi_0 \nonumber\\
\eta & \sim & \alpha_2 \rho + \eta_0 \ ,
\label{param}
\eer
with a non zero tachyon and the constraints
\beq
\frac{1}{2}\alpha_0^2 + \frac{1}{2}\alpha_1^2 - 5\alpha_2^2 = 0 \ , 
\eeq
and $\alpha_i > 0, i=0,1,2$.
With the parametrization (\ref{param}) the metric (\ref{metric}) 
takes the form
\beq
ds =  e^{(\frac{1}{2}\alpha_0+ \frac{1}{2}\alpha_1 - 5\alpha_2)\rho} d\rho^2 +
e^{(\frac{1}{2}\alpha_0-\frac{1}{2} \alpha_1)\rho} dx^{\mu} dx^{\mu} + 
e^{(\frac{1}{2}\alpha_0+ \frac{1}{2}\alpha_1 - \alpha_2)\rho}d\Omega_5^2 \ .
\label{metric1}
\eeq

The large $\rho \sim 1/u^4$ region  corresponds to the IR region of 
the gauge theory.
To analyse the phase of the corresponding gauge theory as a function 
of the gravity parameters $\alpha_i$ we can use the Wilson loop  
as an order parameter. 
Its computation is done as in \cite{Rey:1998ik,Maldacena:1998im}, 
by minimization of the action of a string with worldsheet bounded by the 
loop.
The theory is in a confining phase if the $g_{\mu\mu}$ component of the 
metric (\ref{metric1}) does not vanish anywhere. 
In particular it must not
vanish as $\rho \to \infty$. The reason is that the string
tends to minimize its length by going to the region with the smallest
possible value of $g_{\mu\mu}$, it will stay there for most of the 
geodesic and then will go 
up to the position of the other external charge 
\cite{Brandhuber:1998wl,Gross:1998gk}. 
Therefore the energy
is proportional to the distance of the charges and the string tension is
${\rm Min}(g_{\mu\mu}(\rho))$ up to a numerical
constant. 
On the other hand if there is a zero at the horizon there are two 
possibilities:
either the quark antiquark potential is Coulombic  
\cite{Rey:1998ik,Maldacena:1998im} or the energy is minimized by two
straight strings which can be separated without cost of energy which
corresponds to electric screening. 
In \cite{Minahan:1998af} it was observed that the second possibility is
realized for backgrounds of the form \equ{param}.
Thus, the background 
(\ref{metric1}) corresponds to a phase with electric
screening for $\alpha_0 < \alpha_1$ 
and to a confining phase  for  $\alpha_0 \geq \alpha_1$. 
Using calculations along the lines of \cite{Csaki:1999ln,Minahan:1998af} one
can show the existence of a mass gap in the confining phase.
The radius of the five-sphere is controlled by 
$\alpha_0+\alpha_1 - 2\alpha_2$. When this quantity is negative the 
sphere shrinks to zero size in the IR which signals that the scalars 
are massive and that the theory may be in the same universality class as 
$SU(N)$ Yang-Mills theory.
These solutions at large $\rho$ 
have not been connected to the UV region at small $\rho$ yet. 
With a vanishing tachyon at large $\rho$, $\xi$ and $\eta$ approach a 
constant value and there is a large $\rho$ solution of the form 
$AdS_5 \times S^5$ that signals the IR fixed point discussed above 
\cite{Klebanov:1998af}.

An important issue in the gravity description is the tachyon instability 
\cite{Klebanov:1998db}.
A possible mechanism for removing it employs the RR fields of 
Type 0 theories,
since the effective action contains the terms
\beq
S_{eff} \sim \int \sqrt{G}\left(m^2 e^{-2 \Phi} T^2 + F^2 T^2 \right) \ ,
\label{Tcouple}
\eeq
where $F$ is the RR five form field strength.
The coupling of $T$ and $F$ in (\ref{Tcouple}) can shift the 
effective tachyon mass to be positive.
Assuming the overall scale  of the metric to be $(g_sN)^{1/2}$ results in 
a stability condition $g_s^2 l_s^2  |F|^2 > O(1)$.
Since $F \sim N$ we get $g_sN < O(1)$.
Therefor the stability condition appears to hold for small bare 't Hooft
coupling which is the correct limit to take in order to get to the 
continuum theory.
The same conclusion applies to the cases where the branes are placed at 
singularities which we will study next.

\section{D3 branes in conifold background}

In this section we will study the theory obtained by placing Type 0B D3 branes
at a conifold $\cC$.

\subsection{Field theory}

When placing $N$ D3 branes of Type IIB string theory at a conifold
it is argued \cite{Klebanov:1998sf}
that the resulting worldvolume theory is $\cN=1$ supersymmetric
$SU(N) \times SU(N)$ gauge theory  
with chiral superfields $A_k,k=1,2$ transforming in the $(N,\bar{N})$
representation and $B_l, l=1,2$ transforming in the $(\bar{N},N)$ 
representation.
In the IR the theory flows to a non trivial fixed point with the
superpotential 
$W = \frac{t}{2}\varepsilon^{ij}\varepsilon^{kl}Tr A_iB_kA_jB_l$
becoming exactly marginal.\\

{\bf Electric or magnetic D3 branes}\\

On the worldvolume of $N$ electric
D3 branes of Type 0B string theory at a conifold
we expect a truncation of the fermions, namely, an $SU(N) \times SU(N)$ 
gauge theory  
with complex scalar fields $A_k,k=1,2$ transforming in the $(N,\bar{N})$
representation and $B_l, l=1,2$ transforming in the $(\bar{N},N)$ 
representation.
Classically the theory has $SU(2)\times SU(2)\times U(1)$ global symmetry
which can be viewed as inherited from the $\cN=1$ supersymmetric theory 
upon truncating
the fermionic fields.
One $SU(2)$ acts on $A_k$ and the other $SU(2)$ acts on $B_l$.
The $U(1)$ charges of $A_k$ and $B_l$ are opposite.

With this global symmetry there are several possible bare parameters.
Let us assume the the matrices $A_k,B_l$ are diagonal in some basis.
There are two gauge couplings $g_{YM_1}, g_{YM_2}$, two possible mass terms
$m_1^2 Tr A_k \bar{A}_k,   m_2^2 Tr B_l \bar{B}_l$ and several possible
quartic couplings $g_1 Tr (A_k \bar{A}_k)^2$, 
 $g_2 Tr (A_k \bar{A}_k)(B_l \bar{B}_l)$,  $g_3 Tr (B_l \bar{B}_l)^2$ and squares
of the mass terms.
However, the eigenvalues $a_k,b_l$  of the matrices
$A_k, B_l$ respectively parametrize the positions
of the $N$ D3 branes at points in conifold and are related by the equation
defining the conifold,  $a_k \bar{a_k} = b_l \bar{b_l}$ modulo the $U(1)$
action. This leaves us with one mass parameter $m$ and one quartic 
coupling $g$. 
By symmetry argument one can identify
the bare gauge couplings parameters which we will call $g_{YM}$. 

The Yang-Mills coupling at energy scale $\mu$  
\be
g_{YM}^2(\mu) = \frac{8 \pi^2}{b_1 \log \frac{\mu}{\mu_0} +
\frac{b_2}{2 b^2_1} \log \log \frac{\mu}{\mu_0}} \ , 
\label{beta}
\ee
where $b_1$ and $b_2$ are the one-loop and two-loop beta functions 
coefficients.
The one-loop $\beta$-function coefficient is
$b_1 = 3N$ and the theory is asymptotically free. With the two equal gauge 
couplings the two-loop $\beta$-function coefficient at large $N$ is 
$b_2 = 2N^2$. 
Therefor, unlike
the theory of $N$ electric
D3 branes in flat space, here $\frac{b_2}{2b^2_1} = \frac{1}{9} > 0$.
As we will see later, while the gravity description captures the
asymptotic freedom property of this theory it does not capture
correctly the sign of the  two-loop $\beta$-function.  
The theory has a phase diagram depending on the three parameters $g_{YM},m,g$.
We expect to see in the diagram Coulomb-like phases, confinement phases
and maybe non trivial RG fixed points arising from particular tuning of the 
parameters.

The worldvolume theory of $N$ magnetic
D3 branes of Type 0B string theory is the same as that of $N$ electric ones.\\ 

{\bf Self-dual D3 branes}\\

The theory of $N$ self-dual D3 branes consists of $N$ electric and $N$ magnetic
D3 branes.
The worldvolume theory of $N$ self-dual  
D3 branes at a conifold is a gauge theory with gauge group $SU(N)^4$.
The matter content consists of $8N^2$ complex scalars and $12N^2$ Weyl 
fermions.
Their representation can be deduced as in \cite{Klebanov:1998an}. 
The  $8N^2$ complex scalars transform in the bi-fundamental representations
\beq
2 \left((N,\bar{N},1,1)\oplus (\bar{N},N,1,1)\oplus(1,1,N,\bar{N})\oplus 
(1,1,\bar{N},N)\right) \ ,
\eeq
and the Weyl fermions in
\bea
2(N,1,\bar{N},1)\oplus 2(\bar{N},1,N,1)\oplus 2(1,N,1,\bar{N})\oplus 
2(1,\bar{N},1,N) \nonumber\\
\oplus (N,1,1,\bar{N})\oplus (\bar{N},1,1,N)\oplus (1,N,\bar{N},1)\oplus 
(1,\bar{N},N,1) \ .
\eea
The fermions arise from the open strings stretched between the electric 
and the magnetic branes \cite{Bergman:1997an}.
 
The worldvolume theory is asymptotically free. We will see that the gravity 
picture indicates that for zero vacuum expectation value for the scalars 
the theory flows in the IR to a fixed point.

\subsection{Type 0B gravity description}

In this section we will paramatrize 
the ten dimensional metric in the Einstein frame by 
\bea
ds_E^2 & = & e^{\frac{1}{2} \xi - 3 \lambda + \eta} d\rho^2 +
e^{-\frac{1}{2} \xi} dx^{\mu} dx^{\mu} + 
e^{\frac{1}{2} \xi +\lambda - 3 \eta} (d\psi + \cos(\theta_1) d\phi_1 +
\cos(\theta_2) d\phi_2)^2 + \non
&& e^{\frac{1}{2} \xi - \lambda + \eta}\sum_{i=1,2}
(d\theta_i^2 + \sin(\theta_i)^2 d\phi_i^2) \ ,
\label{metric2}
\eea
with the string metric given by
$ds^2 = e^{\frac{1}{2} \phi} ds_E^2$.

In this parametrization
the effective one dimensional action for radial evolution of the fields
takes the form
\beq  
S=\int d\rho[{1 \over 2} \phi'^2+{1 \over 2} \xi'^2-\lambda'^2+3\eta'^2
-2 \lambda'\eta'+{1\over 4} T'^2-V(\phi,\xi,\lambda,\eta,T)] \ .
\label{effact}
\eeq
The potential $V$ is given by
\beq
V = g(T) e^{{1 \over 2} \phi+{1 \over 2} \xi-3\lambda+ \eta}
-e^{-4\eta}+4e^{-2\lambda}-h(T) e^{-2\xi} \ ,
\label{V}
\eeq
where 
\beq
h(T)=P^2f(T)+{Q^2\over f(T)} \ ,
\label{h}
\eeq
with $f(T) = 1 + T + T^2/2$, and we set the string scale $l_s$ to one. 
The function
$h(T)$ arises from the coupling of the tachyon to electric $Q \neq 0$
and magnetic $P \neq 0$ D3 branes.
Furthermore, the equations of motion of \equ{effact} are supplemented
by a zero energy constraint
\be
{1 \over 2} \phi'^2+{1 \over 2} \xi'^2-\lambda'^2+3\eta'^2
-2 \lambda'\eta'+{1\over 4} T'^2 + V(\phi,\xi,\lambda,\eta,T) = 0 ~~.
\label{zeroenergy}
\ee

We will consider a quartic potential $g(T)$  
for the tachyon field in (\ref{V})
\be
g(T) = \frac{1}{2} T^2 - t T^4  \ .
\label{g}
\ee

\subsubsection{Electric D3 branes}

In the following we will study
the theory of electric threebranes ($P=0$) at a conifold singularity.
We will start by setting 
$g(T) = \frac{1}{2} T^2$ which is the usual mass term for the tachyon. We
will include the quartic coupling (\ref{g}) later. 
The equations of motion following from (\ref{effact})
take the form
\bea
&& \phi'' + \frac{1}{4} T^2 e^{\frac{1}{2} \phi + 
\frac{1}{2} \xi - 3 \lambda + \eta} = 0  \ ,\nonumber \\
&& \xi''  + \frac{1}{4} T^2 e^{\frac{1}{2} \phi + 
\frac{1}{2} \xi - 3 \lambda + \eta} + 2 Q^2 f^{-1} e^{-2 \xi} = 0 \ ,
\nonumber \\
&& \lambda'' + \eta'' + \frac{3}{4} T^2 e^{\frac{1}{2} \phi + 
\frac{1}{2} \xi - 3 \lambda + \eta} + 4 e^{- 2 \lambda} = 0 \ , 
\label{equations} \\
&& -2 \lambda'' + 6 \eta'' + \frac{1}{2} T^2 e^{\frac{1}{2} \phi + 
\frac{1}{2} \xi - 3 \lambda + \eta} + 4 e^{- 4 \eta} = 0 \ , \nonumber \\
&& T'' + 2 T e^{\frac{1}{2} \phi + \frac{1}{2} \xi - 3 \lambda + \eta} 
+ 2 Q^2  \frac{f'}{f^2} e^{- 2 \xi} = 0 \ . \nonumber
\eea

{\bf UV freedom}
\\

In the UV we assume that the tachyon condensenses $T = -1$ and in 
order to find a solution we first solve the equations for massless tachyon
and 
consider the mass term as a small perturbation around the zeroth 
order equations.
The zeroth order solution is just $AdS_5 \times S^5$ with constant
dilaton and takes the form
\be 
T= -1,~~ \phi = \phi_0,~~\xi = \log(2 Q) - y,~~\eta = \frac{1}{4}
\log(4) - \frac{y}{2},~~ \lambda = \frac{1}{2} \log(8/3) - y \ ,
\ee
where 
\be 
\rho = e^{-y} \ .
\label{rho}
\ee
$\rho$ is related to the radial 
variable $u$
by $\rho = u^{-4}$.
In the UV region 
$y \gg 1$ and we find an asymptotic solution to the equations
(\ref{equations}) 
\bea
&& T = -1 + \frac{8}{y} + \frac{156}{y^2}\log(y) + 
\ldots \nonumber \\
&& \phi = \log\left(\frac{2^{13}}{27 Q}\right) - 2 \log(y) + \frac{39}{y}
\log(y)  + \ldots \nonumber \\
&& \xi = \log(2 Q) - y + \frac{1}{y} + \frac{39}{2 y^2}
\log(y)  + \ldots \label{solutions} \\
&& \eta = \frac{1}{4}\log(4) - \frac{y}{2} + \frac{1}{y} + \frac{39}{2 y^2}
\log(y)  + \ldots \nonumber \\
&& \lambda = \frac{1}{2}\log\left(\frac{8}{3}\right) - y + \frac{2}{y} + 
\frac{39}{y^2} \log(y) + \ldots
\nonumber
\eea

The Einstein metric is asymptotic to $AdS_5 \times T^{1,1}$  where
\beq
ds^2_{T^{1,1}} = [\frac{1}{9}(d\psi + \cos(\theta_1) d\phi_1 +
\cos(\theta_2) d\phi_2)^2 + 
\frac{1}{6}\sum_{i=1,2}
(d\theta_i^2 + \sin(\theta_i)^2 d\phi_i^2)] \ ,
\label{t11}
\eeq
which
indicates a UV fixed point.
The effective string coupling vanishes at infinite $u$ and the corresponding
gauge theory has a vanishing coupling in accord with the UV freedom of
the theory.

In fact, it follows from the equations (\ref{equations}) and can be seen
from the solutions (\ref{solutions}) that 
we can identify $\lambda = 2 \eta$ up to a constant term.
With such an identity the
whole metric can be rewritten as a five dimensional metric times the
metric of the $T^{1,1}$ manifold multiplied by a $\rho$ dependent function.
As we will discuss in the next section this is an example
of a  general feature, namely the 
effective gravity solution has the form of a five dimensional space whose form
is universal times a five dimensional Einstein space (with a warp factor).
The relevant equations of $\phi$, $\xi$ and $\eta$ are universal and do
not depend on the background in which we put our D3 branes. 

For D3 branes in flat space 
the numerical values of the two loop beta function in field theory
and gravity do not agree, but the sign matches. 
The sign difference the $b_1$ and $b_2$ coefficients was taken as an evidence
that the
field theory may run to an IR fixed point. 

As in \cite{Klebanov:1998af}, from the form of the dilaton we can 
read off the running of the gauge coupling
\be
Q^{1/2} e^{\frac{1}{2}\phi} \sim 
\frac{1}{\log u - \frac{39}{8} \log \log u}  \ ,
\label{2loop}
\ee
which exhibits a logarithmic running.
However comparing (\ref{2loop}) to (\ref{beta}) we see
that in the gravity description 
$\frac{b_2}{2b^2_1} = -\frac{39}{8} <0$, while the field theory
predicts $\frac{b_2}{2b^2_1} = \frac{1}{9} > 0$. 
Thus, for the conifold model neither the numerical values agrees nor the
sign. 
This, as we will see later, is not special to the conifold model
and generically one cannot trust the sigh of the two-loop $\beta$ function
predicted by gravity.

The sign of $b_2/b^2_1$ is neither necessary nor sufficient to prove the
existence of an IR fixed point in the field theory.
As we will see, the gravity description indicates the existence
of a such fixed point at infinite coupling.

For $u \gg 1$ the effective string coupling 
$e^{\phi} \sim \frac{1}{Q \log^2 u}$.
Therefore the string loops are suppressed and the classical string 
description is applicable.
For large $Q$ the curvature in string units is small too.
However, the $\alpha'$ corrections associated with the Weyl tensor
are of the same order as the leading gravity contribution.
Therefor the classical gravity description is not applicable.
Indeed it does not lead to a quantitatively correct running of the 
gauge coupling.\\

{\bf IR fixed point}
\\

In the region $\rho = e^y \gg 1$ corresponding to the IR
region of the gauge theory we have an asymptotic solution to the 
equations (\ref{equations}) of the form 
\bea
&& T = -\frac{16}{y} + \frac{1}{y^2}(-72 \log(y)  + \ldots) \nonumber \\
&& \phi = \log\left( 2^{-5/2} 27^{-1} Q^{-1} \right) + 2 \log(y) + \frac{1}{y}
(-9 \log(y) + \ldots) \nonumber \\
&& \xi = \frac{1}{2}\log(2 Q^2) + y + \frac{9}{y} + \frac{1}{y^2}
(\frac{81}{2} \log(y) + \ldots ) \label{IRsol} \\
&& \eta = \frac{1}{2}\log(2) + \frac{y}{2} + \frac{1}{y} + \frac{1}{y^2}
(\frac{9}{2} \log(y) +  \ldots ) \nonumber \\
&& \lambda = \frac{1}{2}\log\left(\frac{8}{3}\right) + y + \frac{2}{y} + 
\frac{1}{y^2} (9 \log(y) + \ldots) \nonumber
\eea

In the IR limit $u \rightarrow 0$ the tachyon vanishes and the 
metric in the Einstein frame describes $AdS_5 \times T^{1,1}$.
The effective string coupling is infinite in this limit.
Therefore the solution indicates an IR fixed point at infinite gauge coupling.
The classical string description is not sufficient in order to describe the
fixed point. However, away from the fixed point it can provide a good 
description since $e^{\phi} \sim 1/Q$ which is small for large number of branes.
In the IR the curvature in string units is small and the $\alpha'$ 
corrections are suppressed.
We expect therefor that classical string theory can provide a good description
of the whole flow from the UV to the IR besides the fixed point itself  

In the gauge theory we expect that
a fine tuning of the bare parameters is needed to get to the fixed point.
In the supergravity this fine tuning corresponds to a fine tuning of the
constants of integration of the IR solution to match the UV solution.
In particular we do not expect this fine tuning to be the generic
solution.\\

{\bf Confinement}
\\

To find solutions to (\ref{equations}) in the IR that exhibit
confinement we will consider also the quartic part 
of the tachyon field potential (\ref{g}).
Consider the case when the tachyon condenses at a non zero
value $T_0$ and the tachyon potential (\ref{g}) obeys the 
constraint $g'(T_0)=0$. This implies the tachyon is exactly constant and
does not run. For example one could take $f(T) = 1 + T + T^2/2$ and
$g(T) = T^2/2 - T^4/4$ with $T_0 = -1$ but the precise form
of the tachyon potential is not important. 

In this case
an asymptotic solution for large $\rho$ is given, up to exponentially 
suppressed corrections, by 
\be \phi \sim \phi_0 + \alpha_0 \rho ,~~~
\xi \sim \xi_0 + \alpha_1 \rho ,~~~
\eta \sim \eta_0 + \alpha_2 \rho \ .
\label{confine}
\ee
The zero energy condition (\ref{zeroenergy}) imposes the constraint 
\be
\frac{1}{2}(\alpha_0^2 + \alpha_1^2) - 5 \alpha_2^2=0 \ .
\label{zeroenergyconf}
\ee
The conditions $\alpha_i\ge 0, i=0,1,2$ together with
(\ref{zeroenergyconf}) guarantee that the corrections from exponential
term in the equations of motion (\ref{equations})
are suppressed.

With the parametrization (\ref{confine}) the Einstein frame
metric (\ref{metric2}), 
written in the string frame, takes the form
\beq
ds^2 = e^{(\frac{1}{2}\alpha_0+ \frac{1}{2}\alpha_1 - 5\alpha_2)\rho} 
d\rho^2 + e^{(\frac{1}{2}\alpha_0-\frac{1}{2} \alpha_1)\rho} dx^{\mu} 
dx^{\mu} + e^{(\frac{1}{2}\alpha_0+ \frac{1}{2}\alpha_1 - \alpha_2)\rho} 
ds_{T^{1,1}}^2  \ .
\label{metric3}
\eeq

To analyze the phase of the corresponding gauge theory as a function of the 
gravity parameters $\alpha_i$ we can use the Wilson loop  
as an order parameter. 
The discussion is similar to that in section 2.
The background 
(\ref{metric3}) corresponds to electric screening for $\alpha_0 < \alpha_1$ 
and to a confining phase for  $\alpha_0 \geq \alpha_1$.
The radius of the $T^{1,1}$ is controlled by 
$\alpha_0+\alpha_1 - 2\alpha_2$. When this quantity is negative the 
sphere shrinks to zero size in the IR which signals that the scalars 
are massive and that the theory may be in the same universality class as 
$SU(N)\times SU(N)$ Yang-Mills theory.
Note that although in this case the form of the metric is the same as (\ref{metric1})
when the five-sphere shrinks to zero size, we expect the functions $\alpha_i$
to differ since the corresponding gauge groups are different.

\subsubsection{$(P,Q)$ D3 branes}

{\bf Magnetic D3 Branes}\\

Consider $P$ magnetic D3-branes placed in a conifold singularity.
The discussion will be brief since it is very similar to the electric case. 
We consider the tachyon potential
(\ref{g}) without the quartic term.
Using the definition (\ref{rho})
in the UV region we find the asymptotic solution 
\bea
T&=& -1-{8\over y}+100{\log y\over y^2}+\dots \nonumber \\
\phi&=& \log({{2^{14}}\over {27 P}})-2\log y -25 {{\log y} \over y}+\dots 
\nonumber \\
\xi&=& \log(P) -y+{1\over y}+\dots \\
\lambda&=&{1\over 2} \log({8\over 3}) -y+ {2\over y}+\dots \nonumber \\
\eta&=&{1\over 2} \log 2 -{y\over 2}+ {1\over y}+\dots \nonumber
\eea
From the form of the dilaton we read the running of the gauge coupling 
\be
P^{{1\over 2}}e^{{1\over 2}\phi}\sim {1\over 
{\log u+{25\over 8}\log(\log u)}} \ .
\ee
The gravity solution exhibits the UV freedom property and now 
${b_2\over {2b^2_1}}={25\over 8}$.
The magnetic D3 branes theory is the same as the electric one.
However, we see that the gravity description gives different values for 
$b_2$ in the two cases. 
In fact for a general tachyon potential $g(T)$ 
\be
{b_2\over 2b^2_1} = \left({g'(T_0)\over
g(T_0)} \right)^2-{7\over 8} \ ,
\ee
where $T_0$ is the minimum value of the tachyon, which again indicates 
that the sign of $b_2$ cannot be captured by gravity.

In the IR limit 
we find, using $\rho=e^y$ and $y\gg 1$,  
\bea
T&=& {16\over y}+72{\log y\over y^2}+\dots \nonumber \\
\phi&=&\log({2^{-5/2}\over {27 P}})+2\log y -9 {\log y \over y}+\dots 
\nonumber \\
\xi&=&{1\over 2} \log(2P^2) +y+{9\over y}+\dots \\
\lambda&=& {1\over 2}\log({8\over 3})+y+ {2\over y}+\dots \nonumber \\
\eta&=& {1\over 2}\log 2 +{y\over 2}+ {1\over y}+\dots \nonumber
\eea
This is precisely the fixed point electric D3 branes solution,
except the sign of tachyon. This is
due to the sign change of the tachyon tadpole when
we go from the electric to the magnetic D3-brane 
\cite{Klebanov:1998db}. 

When considering the tachyon potential (\ref{g}) with the quartic part.
we can find, as in the electric case,
solution that exhibit confinement in IR.\\

{\bf Self-Dual D3 Branes}\\

Consider now placing  $N$ self-dual D3-branes at a  conifold 
singularity. We will use the metric ansatz (\ref{metric2}) with 
$h(T)$ given by (\ref{h}). The function
$h(T)$ has an extremum at $T_0=0$ and the tachyon equation is solved by
$T=0$ with a constant dilaton $\phi = \phi_0$ and 
\bea 
\xi&=&\log(2N)+\log\rho, \nonumber \\ 
\lambda&=&{1\over 2}\log({8\over 3})+\log\rho,\\
\eta&=&{1\over 2}\log(2) +{1\over 2}\log\rho \ \nonumber .
\eea
This solution has the form 
$AdS_5 \times T^{1,1}$, which indicates that the IR physics of this 
large $N$ brane system is governed by a fixed point.
Note that unlike the self-dual D3-branes in flat space 
studied in \cite{Klebanov:1998an}, here the 
theory is asymptotically free. 
In particular, the  self-dual D3-branes theory can be obtained as an orbifold
of $\cN=4$ theory by a discrete subgroup of the $Z_4$ center of the R-symmetry group $SU(4)$
\cite{Nekrasov:1999on}.
This is not the case with self-dual D3-branes at a  conifold singularity.

\section{D3 branes in general backgrounds}

What we learned from the previous section is that the supergravity
equations for the fields $\phi$, $T$, $\xi$ and $\eta$ do not depend, 
at least to the order we checked,  on whether
we put D3 branes in a flat background or near a conifold.
In the following we will check this for general backgrounds of the
form $M_4 \times Y_6$ where $M_4$ is flat Minkowski space which
corresponds to the D3 brane worldvolume and $Y_6$ is a non-compact
space which is flat or may have a singularity at the origin.
The space $Y_6$  is  a cone over a five-dimensional compact manifold
$X_5$ which is called the horizon\footnote{This is similar to what 
happens in type IIB theory with the
same background. The supergravity equations yield the same harmonic
function for the D3 branes $f = 1 + Q/r^4$ as in a flat background.
In the type 0 the functions $\xi$ and $\eta$ take over the role of the 
harmonic function $f$ and do not depend on the geometry of the horizon
as well.}.
We will consider solutions that in the UV and the IR take the form
$AdS_5 \times X_5$. For generic values of $r$ the background will
be a fibration of $X_5$ over a five-dimensional space i.e. the
metric of the $X_5$ is multiplied by a radial dependent warp-factor.

We will consider the following ansatz for
the metric and the four-form gauge field:
\bea
ds^2 & = & e^{\frac{1}{2} \phi} ds_E^2 ~~,~~ 
ds_E^2 = e^{-\frac{1}{2} \xi} dx^{\mu} dx^{\mu} + 
         e^{\frac{1}{2} \xi - 5 \eta} d\rho^2 + 
         e^{\frac{1}{2} \xi - \eta} d\Omega_X^2 \non
C_{0123}(r) & = & A(r) ~~,~~ F_{0123r} = A'(r) \non
\phi & = & \phi(r) \ .
\eea

The effective supergravity action is given by:
\bea S & = & {\rm Vol}(M_4) \int dr d\Omega_X \sqrt{g} 
\left( e^{-2\phi} \left[ -2(R + 4 g^{rr} (\phi')^2) + \frac{1}{2} g^{rr} (T')^2
+ \frac{m^2}{2} T^2 ) \right] \right. + \non
&& \left. \frac{1}{2} f(T) g^{00}g^{11}g^{22}g^{33}g^{rr} (A')^2 \right) \non
f(T) & = & 1 + T + \frac{1}{2} T^2 ~~,~~ m^2 = -2 \ .
\eea  
The Einstein equations arising 
from variations with respect to the metric components with $X_5$ indices 
give one part which is proportional to the Ricci tensor plus terms that
are proportional to the $X_5$ metric. 
So in general the metric of the horizon has to obey the following
equation
\be
R_{ij} = c g_{ij} \ ,
\ee
where $i,j$ are $X_5$ indices.
It is easy to see that we can make the choice $c=4$ or in other words
$X_5$ can always be chosen to be an Einstein space. Note that a
different choice of the constant $c$ can
be absorbed by constant shifts in $\xi$, $\eta$ and $\phi$. Well known
examples of Einstein spaces are $S^5$ in the case of a flat background, 
the homogeneous spaces $T^{p,q}$ and del Pezzo surfaces.

In all these cases the equations will be of the form \equ{equations} 
with $\la = 2 \eta$ (up to a constant)
and asymptotic solutions will take the form 
\equ{solutions}, \equ{IRsol} up to possible
constant shifts in $\xi$, $\eta$ and $\phi$. 

This means that in general backgrounds of the form $M_4 \times Y_6$
we will find solutions that
correspond to asymptotically free theories in the UV, and 
there will always exist a solution in the IR which describes a non-trivial 
conformal fixed point at infinite coupling. 
We can also construct in the IR a large class of confining
theories with massgap along the lines of section 3.3. 
One important issue is whether there is a gravity description
of the flow from the UV to the IR. This will not be in general the case, 
as we will discuss in the next section.

Note that in view of the above discussion the two-loop beta
function determined by the gravity description
is independent of the gauge theory we have on
the branes and is universal. 

\section{Discussion}

Besides the Type 0 examples, 
there are by now several gravity solutions discussed in the
literature that may provide a dual description
of confining non supersymmetric gauge theories 
\cite{Witten:1998as,Csaki:1999ln,Kehagias:1999or, Gubser:1999dd, Girardello:1999ca, Armoni:1999ci}.
In all of them the gravity solution cannot provide a quantitative description.

In the following we will discuss the issue of finding
solutions in the effective Type 0 gravity theory that describe qualitatively the
whole RG flow from the  UV to the IR.
To be specific we discuss the case of electric branes in flat space, but in view
of the previous section the discussion is similar for branes at singularities.
On the UV side we found a good candidate solution that exhibits asymptotic
freedom which has a description  in terms of classical string theory.
Using numerical methods we tried to continue the known asymptotic solution
in the UV to the IR.
In the case of zero quartic coupling $t=0$ we found that the tachyon goes
from $T_0 = -1$
through an oscillatory region and settles down at zero. Unfortunately
the numerical approximation fails at that point and one is not able
to trust the numerical method and go beyond.

The other possibility is to start with one of the IR solutions (the IR fixed
point or the confining solutions) and proceed to the UV. 
Whereas starting from the IR fixed point one runs into 
similar numerical problems as for the UV solution, 
starting with a confining solution
in the IR, which is the most generic situation in field theory, produced
the most promising results. 
In this case we chose $t=1/4$ such that
$g'(T_0) = 0$ and the tachyon does not run and is constant
$T_0 = -1$ all along the flow. In this situation we were able to find
solutions connecting the IR to the UV but these solutions depend strongly
on the integration constants $\alpha_i, \phi_0, \xi_0, \eta_0$. In some
cases $\phi,\xi,\eta$ settle down at constant values in the UV which 
corresponds to a space time with geometry ${\bf R}^+ \times {\bf R}^4 \times
S^5$. The field theory interpretation of such a solution in the UV is missing.
In other cases one or several of the functions diverge, 
but we were not able to identify a solution
in the UV that asymptotes $AdS_5 \times S^5$ and behaves like the UV solution.
Still, the good behaviour of the solutions we found gives us some 
confidence that such a solution exists, 
but the integration constants have to be chosen carefully.
A generic confining solution in the IR in general does not connect to
a UV solution with asymptotic freedom. To further pursue this direction
requires a better
understanding of the relation between the field theory parameters and
the integration constants.

\section*{Acknowledgement}

We would like to thank K. Jansen for discussions.
M.A. is supported by the John Bell scholarship from the world laboratory.

\newpage

\bibliography{type0}

\begingroup\raggedright\begin{thebibliography}{10}

\bibitem{Dixon:1986ba}
L.~Dixon and J.~Harvey, ``String Theories in Ten Dimensions Without Space-Time
  Supersymmetry,'' {\em Nucl. Phys.} {\bf B274} (1986) 93.

\bibitem{Seiberg:1986ss}
N.~Seiberg and E.~Witten, ``Spin Structures in String Theory,'' {\em Nucl.
  Phys.} {\bf B276} (1986) 272.

\bibitem{Polyakov:1998ju}
A.~M. Polyakov, ``The Wall of the cave,''
  \href{http://xxx.lanl.gov/abs/hep-th/9809057}{{\tt hep-th/9809057}}.

\bibitem{Klebanov:1998db}
I.~R. Klebanov and A.~Tseytlin, ``D-Branes and Dual Gauge Theories in Type 0
  Strings,'' \href{http://xxx.lanl.gov/abs/hep-th/9811035}{{\tt
  hep-th/9811035}}.

\bibitem{Klebanov:1998af}
I.~R. Klebanov and A.~Tseytlin, ``Asymptotic Freedom and Infrared Behavior in
  the Type 0 String Approach to Gauge Theory,''
  \href{http://xxx.lanl.gov/abs/hep-th/9812089}{{\tt hep-th/9812089}}.

\bibitem{Minahan:1998gm}
J.~A. Minahan, ``Glueball Mass Spectra and Other Issues for Supergravity Duals
  of QCD Models,'' \href{http://xxx.lanl.gov/abs/hep-th/9811156}{{\tt
  hep-th/9811156}}.

\bibitem{Minahan:1998af}
J.~A. Minahan, ``Asymptotic Freedom and Confinement from Type 0 String
  Theory,'' \href{http://xxx.lanl.gov/abs/hep-th/9902074}{{\tt
  hep-th/9902074}}.

\bibitem{Baier:1986ts}
R.~Baier, R.~V. Gavai, and C.~Lang, ``Tricritical Structure In The Adjoint
  Higgs Model?,'' {\em Phys. Lett.} {\bf B172} (1986) 387.

\bibitem{Baier:1987ts}
R.~Baier and H.~J. Reusch, ``The SU(2) Adjoint Higgs Model: Phase Structure By
  Mean Field Approximations,'' {\em Nucl. Phys.} {\bf B285} (1987) 535.

\bibitem{Baier:1988tr}
R.~Baier, C.~Lang, and H.~J. Reusch, ``The Renormalization Flow In The Adjoint
  SU(2) Lattice Higgs Model,'' {\em Nucl. Phys.} {\bf B305} (1988) 396.

\bibitem{Witten:1998as}
E.~Witten, ``Anti-de Sitter Space, Thermal Phase Transition, And Confinement in
  Gauge Theories,'' {\em Adv.Theor.Math.Phys.} {\bf 2} (1998) 505--532,
  \href{http://xxx.lanl.gov/abs/hep-th/9803131}{{\tt hep-th/9803131}}.

\bibitem{Csaki:1999ln}
C.~Csaki, Y.~Oz, J.~Russo, and J.~Terning, ``Large N QCD from Rotating
  Branes,'' {\em Phys. Rev.} {\bf D59} (1999) 065008,
  \href{http://xxx.lanl.gov/abs/hep-th/9810186}{{\tt hep-th/9810186}}.

\bibitem{Rey:1998ik}
S.-J. Rey and J.~Yee, ``Macroscopic strings as heavy quarks in large N gauge
  theory and anti-de Sitter supergravity,''
  \href{http://xxx.lanl.gov/abs/hep-th/9803001}{{\tt hep-th/9803001}}.

\bibitem{Maldacena:1998im}
J.~Maldacena, ``Wilson loops in large N field theories,'' {\em Phys. Rev.
  Lett.} {\bf 80} (1998) 4859,
  \href{http://xxx.lanl.gov/abs/hep-th/9803002}{{\tt hep-th/9803002}}.

\bibitem{Brandhuber:1998wl}
A.~Brandhuber, N.~Itzhaki, J.~Sonnenschein, and S.~Yankielowicz, ``Wilson
  loops, confinement, and phase transitions in large N gauge theories from
  supergravity,'' {\em JHEP} {\bf 06} (1998) 001,
  \href{http://xxx.lanl.gov/abs/hep-th/9803263}{{\tt hep-th/9803263}}.

\bibitem{Gross:1998gk}
D.~J. Gross and H.~Ooguri, ``Aspects of large N gauge theory dynamics as seen
  by string theory,'' {\em Phys. Rev.} {\bf D58} (1998) 106002,
  \href{http://xxx.lanl.gov/abs/hep-th/9805129}{{\tt hep-th/9805129}}.

\bibitem{Klebanov:1998sf}
I.~R. Klebanov and E.~Witten, ``Superconformal field theory on three-branes at
  a Calabi-Yau singularity,'' {\em Nucl. Phys.} {\bf B536} (1998) 199,
  \href{http://xxx.lanl.gov/abs/hep-th/9807080}{{\tt hep-th/9807080}}.

\bibitem{Klebanov:1998an}
I.~R. Klebanov and A.~Tseytlin, ``A Non-supersymmetric Large N CFT from Type 0
  String theory,'' \href{http://xxx.lanl.gov/abs/hep-th/9901101}{{\tt
  hep-th/9901101}}.

\bibitem{Bergman:1997an}
O.~Bergman and M.~R. Gaberdiel, ``A Non-Supersymmetric Open String Theory and
  S-Duality,'' {\em Nucl.Phys.} {\bf B499} (1997) 183,
  \href{http://xxx.lanl.gov/abs/hep-th/9701137}{{\tt hep-th/9701137}}.

\bibitem{Nekrasov:1999on}
N.~Nekrasov and S.~L. Shatashvili, ``On Non-Supersymmetric CFT in Four
  Dimensions,'' \href{http://xxx.lanl.gov/abs/hep-th/9902110}{{\tt
  hep-th/9902110}}.

\bibitem{Kehagias:1999or}
A.~Kehagias and K.~Sfetsos, ``On running couplings in gauge theories from type
  IIB supergravity,'' \href{http://xxx.lanl.gov/abs/hep-th/9902125}{{\tt
  hep-th/9902125}}.

\bibitem{Gubser:1999dd}
S.~S. Gubser, ``Dilaton driven confinement,''
  \href{http://xxx.lanl.gov/abs/hep-th/9902155}{{\tt hep-th/9902155}}.

\bibitem{Girardello:1999ca}
L.~Girardello, M.~Petrinia, M.~Porrati, and A.~Zaffaroni, ``Confinement and
  Condensates Without Fine Tuning in Supergravity Duals of Gauge Theories,''
  \href{http://xxx.lanl.gov/abs/hep-th/9903026}{{\tt hep-th/9903026}}.

\bibitem{Armoni:1999ci}
A.~Armoni, E.~Fuchs, and J.~Sonnenschein, ``Confinement in 4D Yang-Mills
  Theories from Non-Critical Type 0 String Theory,''
  \href{http://xxx.lanl.gov/abs/hep-th/9903090}{{\tt hep-th/9903090}}.

\end{thebibliography}\endgroup
\bibliographystyle{ssg}

\end{document}